\documentclass[aps,prd,a4paper,twocolumn,showpacs,nofootinbib]{revtex4-1}

\usepackage{amssymb,amsmath,latexsym,mathrsfs}
\usepackage{graphicx,subfigure}
\usepackage{epsfig}
\usepackage{varioref,xr-hyper}
\usepackage{color}

\newcommand{\Mpc}{{\rm ~Mpc}}
\newcommand{\nnu}{N_{\rm {\nu}}}
\newcommand{\nnus}{N_{\rm {\nu}}^S}
\newcommand{\neff}{N_{\rm {eff}}}
\newcommand{\cvis}{c_{\rm vis}^2}
\newcommand{\ceff}{c_{\rm eff}^2}

\begin{document}

\title{The Case for Dark Radiation}

\author{Maria Archidiacono$^{a}$}
\author{Erminia Calabrese$^{a}$}
\author{Alessandro Melchiorri$^{a}$}

\affiliation{$^a$ Physics Department and INFN, Universit\`a di Roma ``La Sapienza'', Ple Aldo Moro 2, 00185, Rome, Italy}

\begin{abstract}
Combined analyses of recent cosmological data are showing interesting hints for the presence of an extra relativistic component, coined Dark Radiation.  Here we perform a new search for Dark Radiation, parametrizing it with an effective number of relativistic degrees of freedom parameter, $\neff$. We show that the cosmological data we considered are clearly suggesting the presence for an extra relativistic component with $\neff=4.08_{-0.68}^{+0.71}$ at $95 \%$ c.l.. Performing an analysis on Dark Radiation sound speed $c_{\rm eff}$ and viscosity $c_{\rm vis}$ parameters, we found  $\ceff=0.312\pm0.026$ and $\cvis=0.29_{-0.16}^{+0.21}$ at $95 \%$ c.l., consistent with the expectations of a relativistic free streaming component ($\ceff$=$\cvis$=$1/3$). Assuming the presence of $3$ relativistic neutrinos we constrain the
extra relativistic component with $\nnus=1.10_{-0.72}^{+0.79}$ and $\ceff=0.24_{-0.13}^{+0.08}$ at $95 \%$ c.l. while $\cvis$ results as unconstrained. Assuming a massive neutrino component we obtain
further indications for Dark Radiation with $\nnus=1.12_{-0.74}^{+0.86}$ at $95 \%$ c.l. .
\end{abstract}
\maketitle

\section{Introduction}

Since almost a decade, observations from Cosmic Microwave Background (CMB hereafter) satellite,
balloon-borne and ground based experiments (\cite{wmap7}, \cite{act}, \cite{acbar}, \cite{spt}),
galaxy redshift surveys \cite{red} and luminosity distance measurements,
are fully confirming the theoretical predictions  of the standard $\Lambda$CDM
cosmological model. This not only permits to place stringent constraints
on the parameters of the model but can be fruitfully used to constrain non
standard physics at the fundamental level, such as classes of elementary particle
models predicting a different radiation content in the Universe.

One of the major theoretical predictions of the standard scenario is the existence of
a relativistic energy component ( see e.g. \cite{kolb}), beside CMB photons,
with a current energy density given by :

\begin{equation}
\rho_{rad}=\Big [1+{7 \over 8} \big({4 \over 11}\big )^{4/3} \neff \Big ]\rho{_\gamma} \ ,
\end{equation}

where $\rho_{\gamma}$ is the energy density of the CMB photons
background at temperature $T_{\gamma}=2.728K$ and $\neff$ is
in principle a free parameter, defined as the effective number of relativistic degrees
of freedom.  Assuming standard electroweak interactions, three active
massless neutrinos and including the (small) effect of neutrino
flavour oscillations the expected value is $\neff=3.046$ with a deviation from
 $\neff=3$ that takes into account effects from the non-instantaneous neutrino
decoupling from the primordial photon-baryon plasma (see e.g. \cite{mangano3046}).

In recent years, thanks to the continuous experimental advancements, the value of $\neff$ has been
increasingly constrained from cosmology (\cite{bowen}, 
\cite{seljak06}, \cite{cirelli}, \cite{mangano07}, \cite{ichikawa07},
\cite{wmap7}, \cite{hamann10}, \cite{giusarma11}, \cite{krauss}, \cite{reid}, \cite{riess},
\cite{knox11}), ruling out $\neff=0$ at high significance.

However, especially after the new ACT \cite{act} and SPT \cite{spt} CMB results,
the data seem to suggest values higher than the "standard" one,
 with $\neff \sim4-5$ (see e.g. \cite{hamann10}, \cite{giusarma11}, \cite{riess}, \cite{knox11}, 
 \cite{zahn}) in tension with the expected standard value at about two standard deviations.

The number of relativistic degrees of freedom obviously depends on the
decoupling process of the neutrino background from the primordial plasma.
However, a value of $\neff=4$ is difficult to explain in the three neutrino
framework since non-standard neutrino decoupling is expected to maximally
increase this value up to $\neff \sim 3.12$ (see e.g. \cite{mangano06}).
A possible explanation could be the existence of a fourth (or fifth) sterile neutrino.
The hypothesis of extra neutrino flavour is interesting since recent results
from short-baseline neutrino oscillation data from LSND \cite{lsnd} and
MiniBooNE \cite{minibun} experiments are consistent with a possible
fourth (or fifth) sterile neutrino specie (see \cite{hamann10,giusarma11}
and references therein). Moreover, a larger value for $\neff \sim 4$
could arise from a completely different physics, related to
axions (see e.g. \cite{axions}), gravity waves (\cite{gw}),
decaying particles (see e.g. \cite{decay}), extra dimensions \cite{extra,Hebecker:2001nv}
and dark energy (see e.g. \cite{ede} and references therein).

As a matter of fact, any physical mechanism able to produce extra "dark" radiation
produces the same effects on the background expansion 
of additional neutrinos, yielding a larger value for $\neff$ from observations.

Since there is a large number of models that could enhance $\neff$ it is clearly
important to investigate the possible ways to discriminate among them. 
If Dark Radiation is made of relativistic particles as sterile neutrinos it should
behave as neutrinos also from the point of view of perturbation theory, i.e. 
if we consider the set of equations that describes perturbations in massless neutrino
(following the definition presented in \cite{gdm}):

\begin{align}
&\dot{\delta}_{\nu} =  \frac{\dot{a}}{a} (1-3 \ceff) \left(\delta_{\nu}+3 \frac{\dot{a}}{a}\frac{q_{\nu}}{k}\right)-k \left(q_{\nu}+\frac{2}{3k} \dot{h}\right), \\
&\dot{q}_{\nu}  =  k \ceff \left(\delta_{\nu}+3 \frac{\dot{a}}{a}\frac{q_{\nu}}{k}\right)- \frac{\dot{a}}{a}q_{\nu}- \frac{2}{3} k \pi_{\nu}, \\
&\dot{\pi}_{\nu} =  3 \cvis \left(\frac{2}{5} q_{\nu} + \frac{8}{15} \sigma\right)-\frac{3}{5} k F_{\nu,3}, \\
&\frac{2l+1}{k}\dot{F}_{\nu,l} -l F_{\nu,l-1} =  - (l+1) F_{\nu,l+1},\ l \geq 3 \ ,
\end{align}

\noindent it should have an effective sound speed $c_{\rm eff}$ and a viscosity speed 
$c_{\rm vis}$ such that $\ceff=\cvis=1/3$. Free streaming of relativistic neutrino will indeed produce
anisotropies in the neutrino background yielding a value of $\cvis=1/3$ while
a smaller value would indicate possible non standard interactions (see e.g. \cite{couplings}).
A value of $c_{\rm vis}$ different from zero, as expected in the standard
scenario, has been detected in \cite{trotta} and confirmed in subsequent papers
\cite{dopo}. More recently, the analysis of \cite{zahn} confirmed the presence of anisotropies
from current cosmological data but also suggested the presence of a lower value for
the effective sound speed with $\ceff=1/3$ ruled out at more than two standard deviations.

Given the current situation and the experimental hints for $\neff \sim4$ is therefore timely to 
perform a new analysis for $\neff$ (and the perturbation parameters $\ceff$ and $\cvis$) 
with the most recent cosmological data. This is the kind of analysis we perform in this paper,
organizing our work as follows: in Sec. II we describe the data and the data analysis
method adopted.
We present our results in the first two subsections of Sec. III,
depending on two adopted different parametrizations for the Dark Radiation.
Moreover a model independent analysis is also discussed in the last subsection of Sec. III.
Finally we conclude in Sec. IV.

\section{Analysis Method}

We perform a COSMOMC \cite{Lewis:2002ah} analysis combining the following CMB
datasets: WMAP7 \cite{wmap7}, ACBAR \cite{acbar}, ACT \cite{act}, and SPT \cite{spt},
and we analyze datasets using out to $l_{\rm max}=3000$.  We also include
information on dark matter clustering from the galaxy power spectrum
extracted from the SDSS-DR7 luminous red galaxy sample
\cite{red}. Finally, we impose a prior on the Hubble parameter based
on the last Hubble Space Telescope observations \cite{hst}.

The analysis method we adopt is based on the publicly available Monte Carlo
Markov Chain package \texttt{cosmomc} \cite{Lewis:2002ah} with a convergence
diagnostic done through the Gelman and Rubin statistic.
We sample the following six-dimensional standard set of cosmological parameters,
adopting flat priors on them: the baryon and cold dark matter densities
$\Omega_{\rm b}$ and $\Omega_{\rm c}$, the ratio of the sound horizon to the angular
diameter distance at decoupling $\theta$, the optical depth to reionization $\tau$,
the scalar spectral index $n_S$, and the overall normalization of the
spectrum $A_S$ at $k=0.002\Mpc^{-1}$. We consider purely adiabatic initial conditions and
we impose spatial flatness.
We vary the effective number of relativistic degrees of freedom $\neff$, the
effective sound speed $\ceff$, and the viscosity parameter $\cvis$.
In some cases, we consider only variations in the {\it extra} dark radiation
component $\nnus=\neff-3.046$, varying the perturbation parameters $c_{\rm vis}$ and
$c_{\rm eff}$ only for this extra component and assuming $\ceff=\cvis=1/3$ for the 
standard $3$ neutrino component.

In our analysis we always fix the primordial Helium abundance to the observed value $Y_p=0.24$. 
This  procedure is different from the one adopted, for example, in \cite{spt}, where
the $Y_p$ parameter is varied assuming Big Bang Nucleosynthesis for each values of
$\neff$ and $\Omega_{\rm b}$ in the chain. Since the cosmological epoch and the energy scales 
probed by BBN are dramatically different from the ones probed by CMB and large scale
structure we prefer to do not assume standard BBN in our analysis and to leave
the primordial Helium abundance as fixed to a value consistent with current observations. 

We account for foregrounds contributions including three extra amplitudes:
the SZ amplitude $A_{SZ}$, the amplitude of clustered point sources $A_C$,
and the amplitude of Poisson distributed point sources $A_P$. 
We marginalize the contribution from point sources only for the ACT and SPT data, based on the templates provided by \cite{spt}. We quote only one joint amplitude parameter for each component (clustered and Poisson distributed). Instead, the SZ amplitude is obtained fitting the WMAP data with the WMAP own template, while for SPT and ACT it is calculated using the \cite{Trac:2010sp} SZ template at 148 GHz. Again, this is different from the analysis performed
in \cite{spt} where no SZ contribution was considered for the WMAP data.

\section{Results}

As stated in the previous section, we perform two different analyses.
In the first analysis we vary the amplitude of the 
whole relativistic contribution changing $\neff$ and the corresponding
perturbation parameters $\cvis$ and $\ceff$. In the second analysis we
assume the existence of a standard neutrino background and vary only
the extra component $\nnus=\neff-3.046$ considering only in this
extra component the variations in $\cvis$ and $\ceff$.

\subsection{Varying the number of relativistic degrees of freedom $\neff$ .}

\begin{figure*}[]
\includegraphics[scale=0.31]{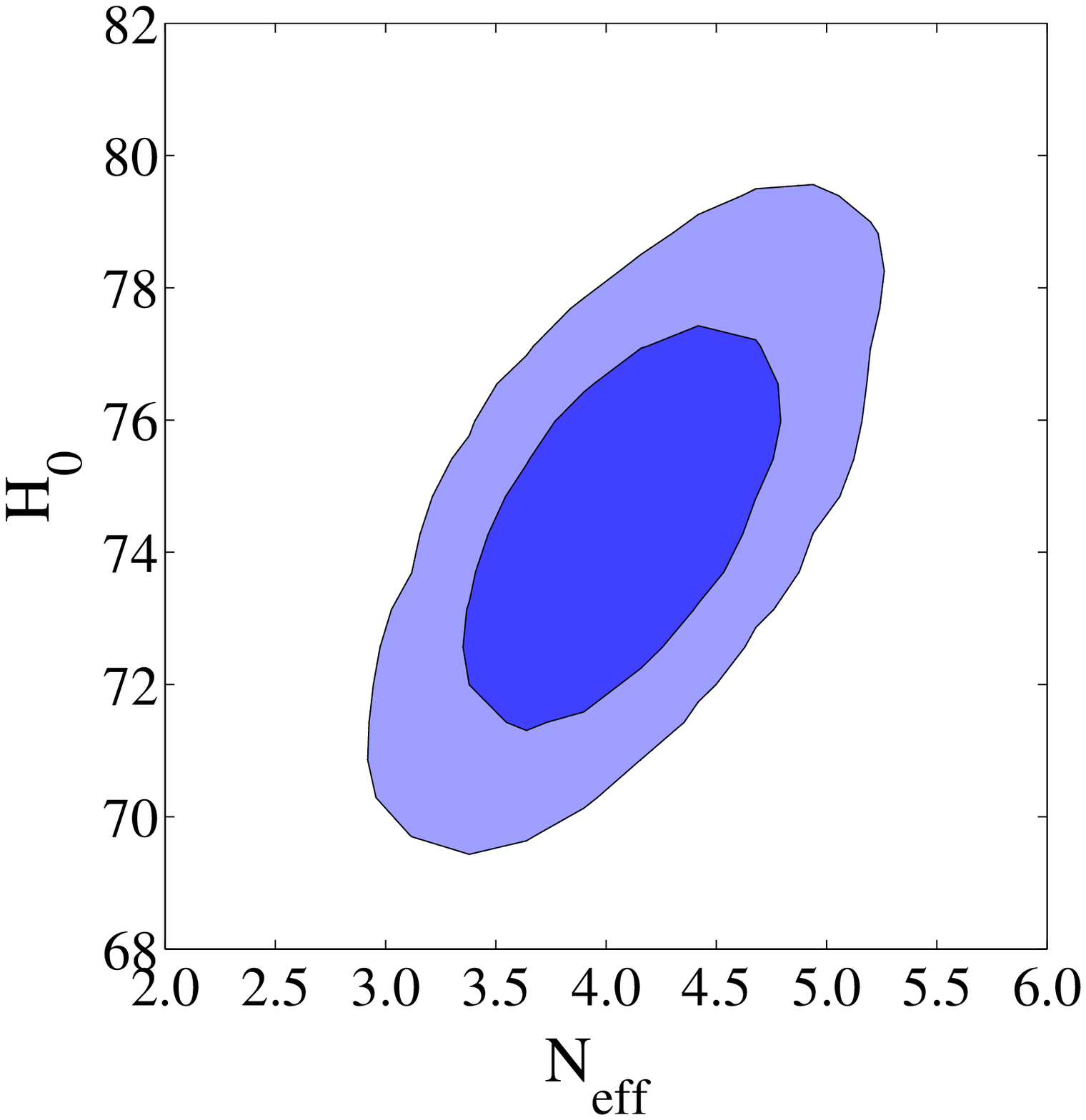}
\includegraphics[scale=0.31]{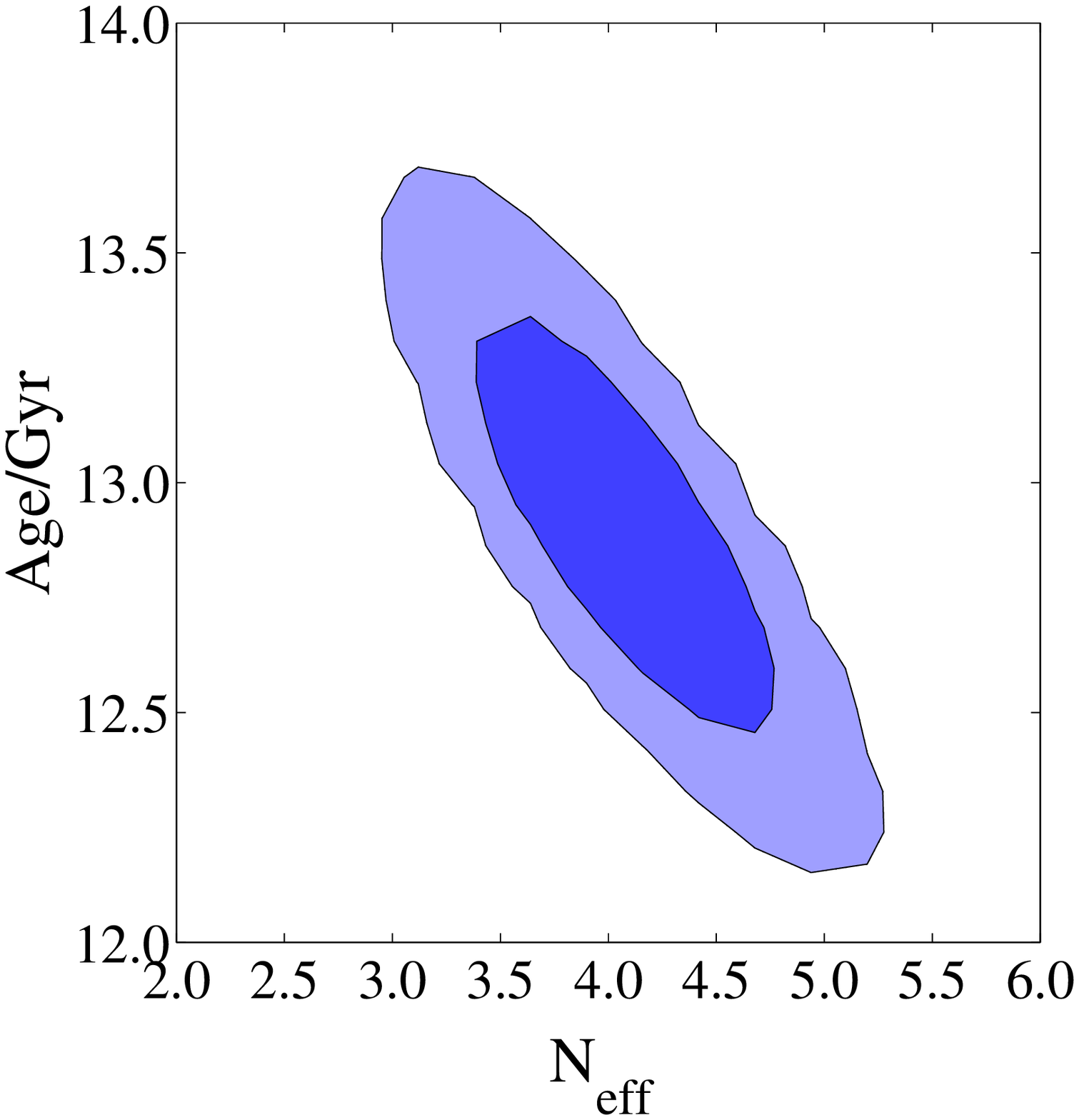}
\includegraphics[scale=0.31]{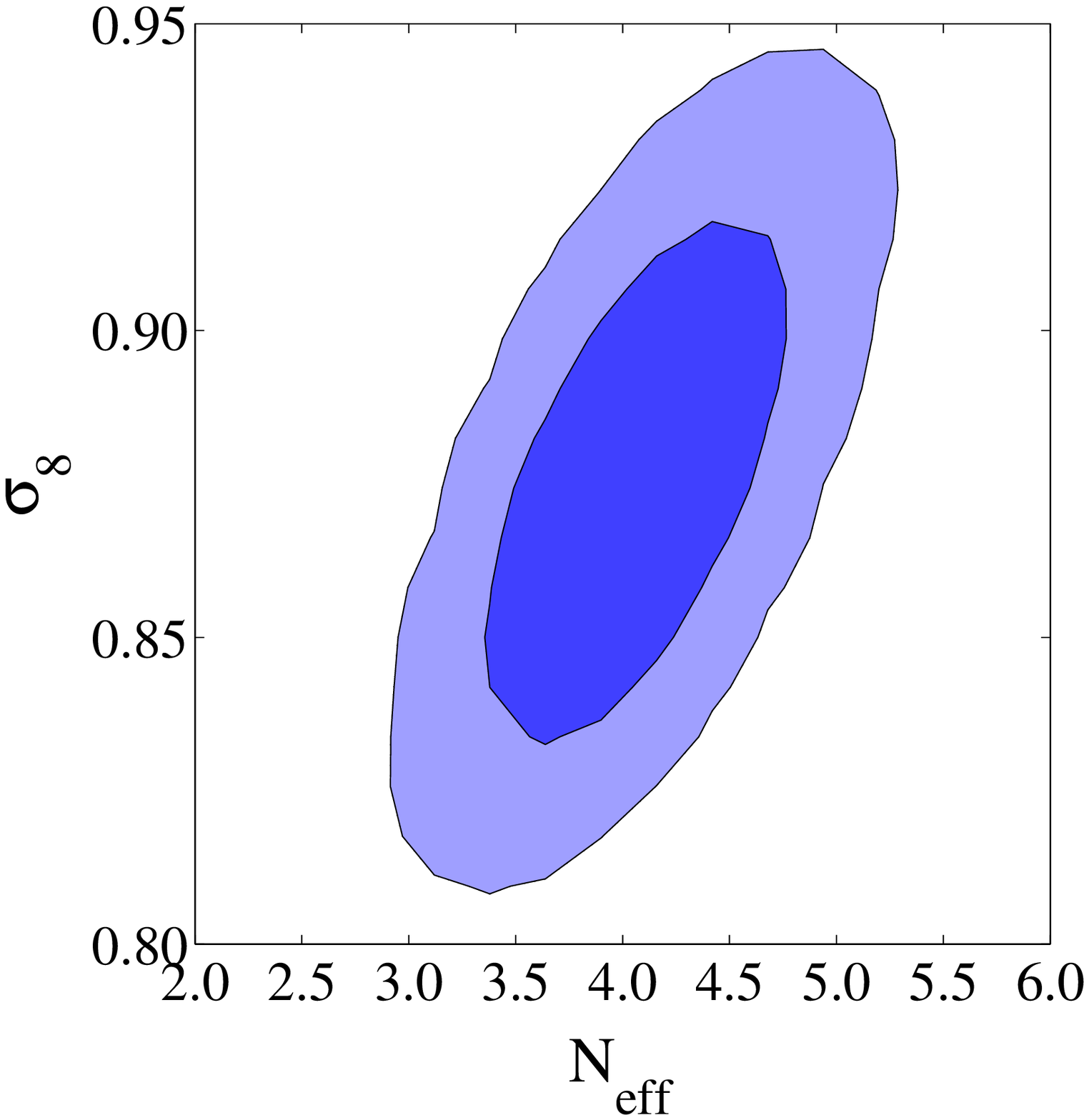}
\caption{$68\%$ and $95\%$ c.l. constraints for the
degeneracy between $\neff$ and the Hubble constant $H_0$, the age of the universe
$t_0$, and the amplitude of mass fluctuations $\sigma_8$.}
\label{degenerazioni}
\end{figure*}

In Table \ref{standard} we report the constraints on the cosmological parameters
varying $\neff$ with and without variations in perturbation theory. 
We consider two cases: first we run our analysis
fixing the perturbation parameters to the standard values, i.e. $\ceff=\cvis=1/3$,
then we let those parameters to vary freely.

\begin{table}[h!]
\begin{center}
\begin{tabular}{|l|c|c|}
\hline
\hline
$\Omega_b h^2$ & $0.02229 \pm 0.00038$ & $0.02206 \pm 0.00081$\\
$\Omega_c h^2$ & $0.1333 \pm 0.0086$ & $0.1313 \pm 0.0094$\\
$\tau$ & $0.082 \pm 0.012$ & $0.083 \pm 0.014$\\
$H_0$ & $74.3 \pm 2.2$ & $74.2 \pm 2.1$\\
$n_s$ & $0.977 \pm 0.011$ & $0.972 \pm 0.021$\\
$log(10^{10} A_s)$ & $3.195 \pm 0.035$ & $3.196 \pm 0.035$\\
$A_{SZ}$ & $< 1.2$ & $< 1.4$\\
$A_C [{\rm \mu K^2}]$ & $<14.3$ & $< 14.6$\\
$A_P [{\rm \mu K^2}]$ & $<25.2$ & $< 24.7$\\
\hline
$\neff$ & $4.08^{+0.18 +0.71}_{-0.18 -0.68}$ & $3.89^{+0.19 +0.70}_{-0.19 -0.70}$\\
$\ceff$ & $1/3$ & $0.312^{+0.008 +0.026}_{-0.007 -0.026}$\\
$\cvis$ & $1/3$ & $0.29^{+0.04 +0.21}_{-0.06 -0.16} $\\
\hline
$\chi^2_{min}$ & $7594.2$ & $7591.5$\\
\hline
\hline
\end{tabular}
\caption{MCMC estimation of the cosmological parameters
assuming $\neff$ relativistic neutrinos.
Upper bounds at $95 \%$ c.l. are reported for foregrounds parameters.
We quote the one-dimensional marginalized $68\%$ and $95\%$ c.l. for the neutrino parameters.}
\label{standard}
\end{center}
\end{table}

As we can see from the results in the left column of Table \ref{standard}, the
WMAP7+ACT+SPT+DR7+H0 analysis
is clearly suggesting the presence for Dark Radiation with $\neff = 4.08_{-0.68}^{+0.71}$
at $95 \%$ c.l.. When considering variations in the perturbation parameters
(right column) the constraint is somewhat shifted
towards smaller values with $\neff = 3.89^{+0.70}_{-0.70}$. The constraint
on the sound speed, $\ceff = 0.312\pm0.026$ is fully consistent with
the expectations of a free streaming component. Anisotropies in the neutrino
background are detected at high statistical significance with
$\cvis=0.29^{+0.21}_{-0.16}$ improving previous constraints presented in
\cite{trotta}.

It is interesting to consider the possible degeneracies between $\neff$ and other
"indirect" (i.e. not considered as primary parameters in MCMC runs)
model parameters. In Figure \ref{degenerazioni} we therefore plot the 2D likelihood
constraints on $\neff$ versus the Hubble constant $H_0$, the age of the universe
$t_0$ and the amplitude of r.m.s. mass fluctuations on spheres of $8 {\rm Mpc} h^{-1}$,
$\sigma_8$.

As we can see from the three panels in the figure, there is a clear degeneracy between
$\neff$ and those three parameters. Namely, an extra radiation component will bring
the cosmological constraints (respect to the standard $3$ neutrino case) to
higher values of the Hubble constant and of $\sigma_8$ and to lower values of the
age of the universe $t_0$. These degeneracies have been already discussed in the literature
(see e.g. \cite{nuage}) and could be useful to estimate the effect of additional
datasets on our result. The $3 \%$ determination of the Hubble constant from
the analysis of \cite{riess} plays a key role in our analysis in shifting the
constraints towards larger values of $\neff$. If future analyses will point
towards lower values of the Hubble constant, this will make the standard
$3$ neutrino case more consistent with observations.
If future observations will point towards values of the age of the universe
significantly larger than $13$ Gyrs, this will be against an extra dark radiation
component, since it prefers $t_0\sim 12.5 {\rm Gyrs}$.
Clearly, adding cluster mass function data as presented in \cite{clusters}
and that points towards lower values of $\sigma_8$ renders the standard $\neff=3.046$
case more consistent with observations. A future and precise determination
of $\sigma_8$ from clusters or Lyman-$\alpha$ surveys could be crucial in ruling out
dark radiation.

\subsection{Varying only the excess in the relativistic component $\nnus$ and assuming $3$
standard neutrinos.}

\begin{figure*}[]
\includegraphics[scale=0.315]{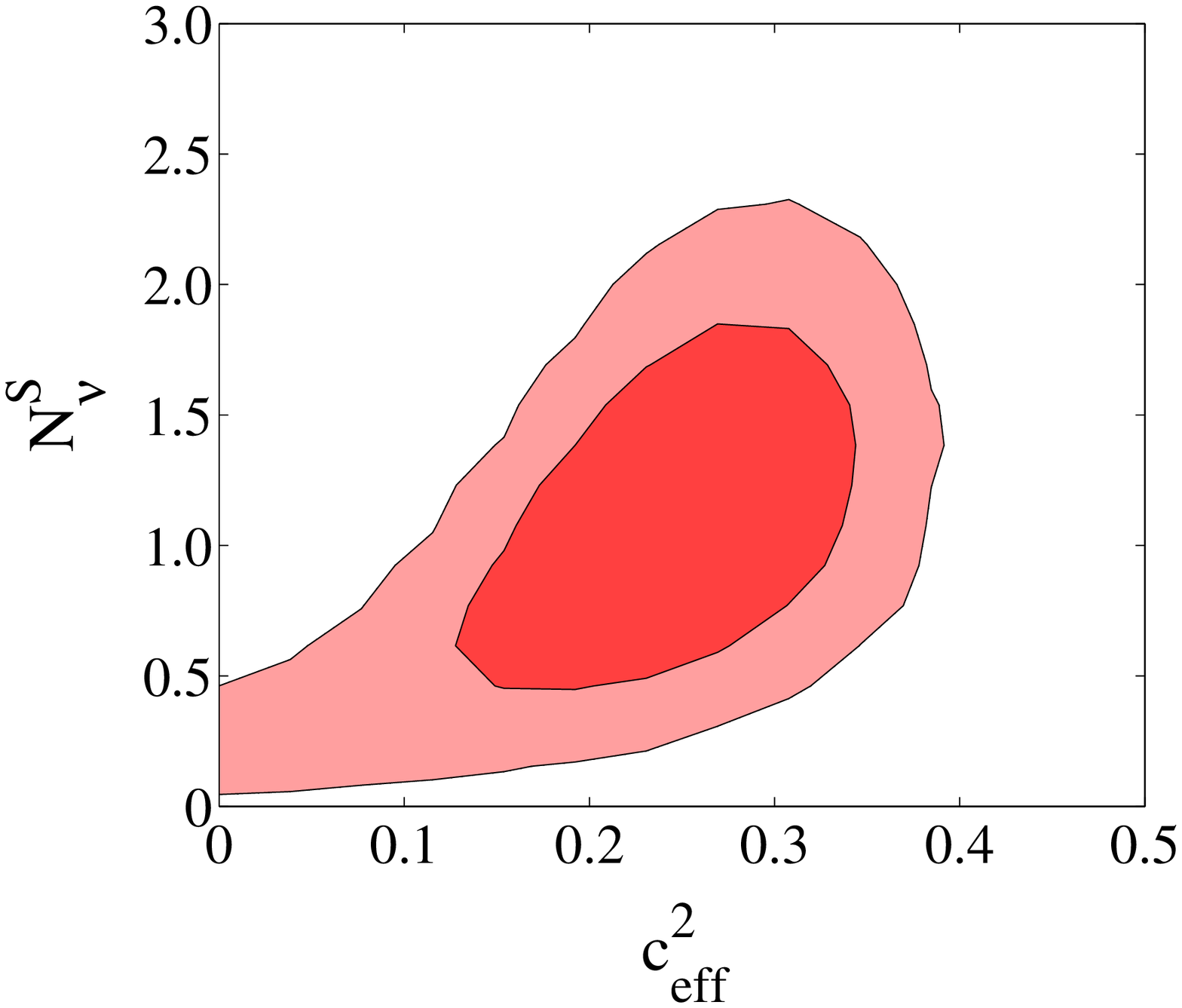}
\includegraphics[scale=0.315]{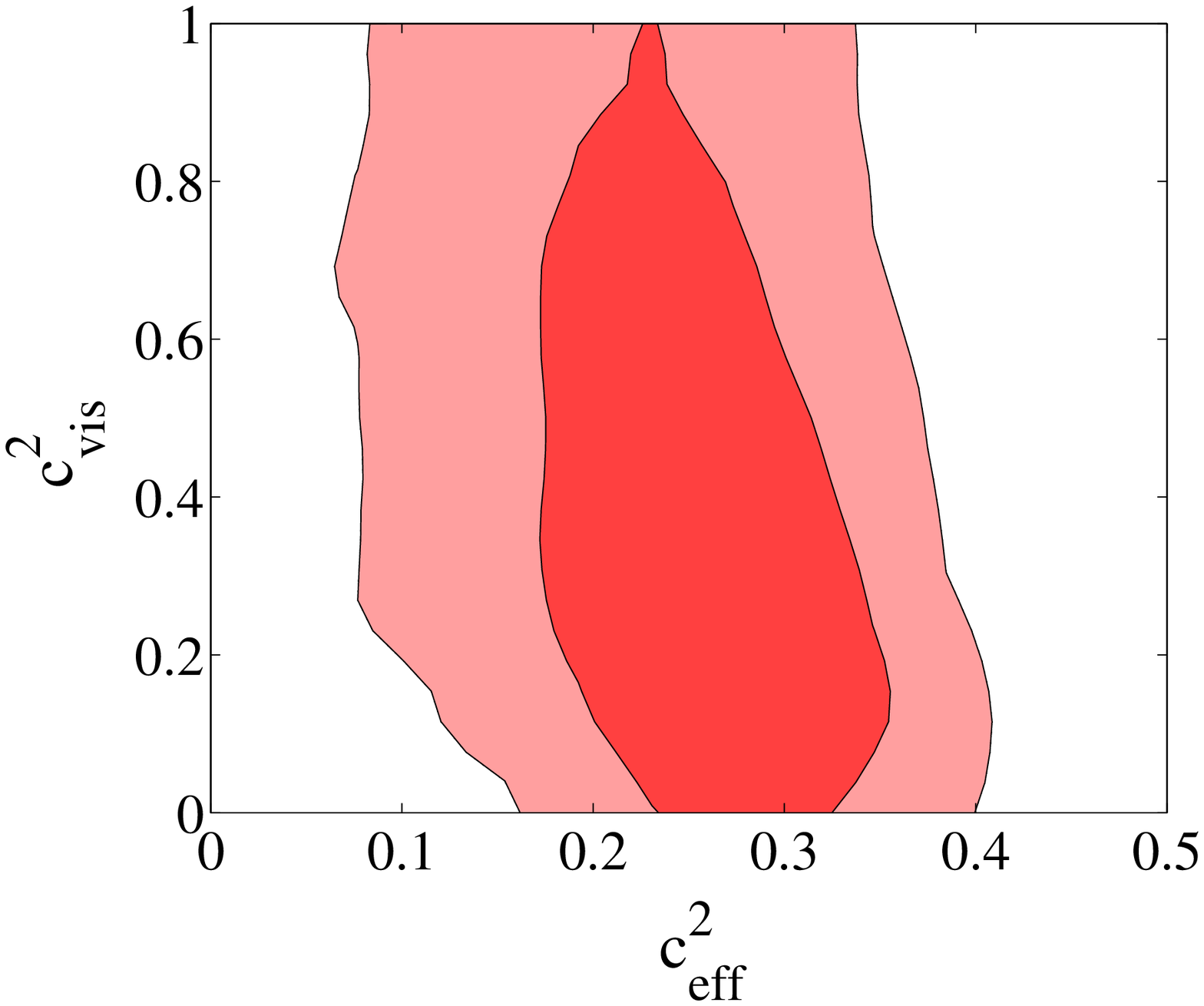}
\includegraphics[scale=0.315]{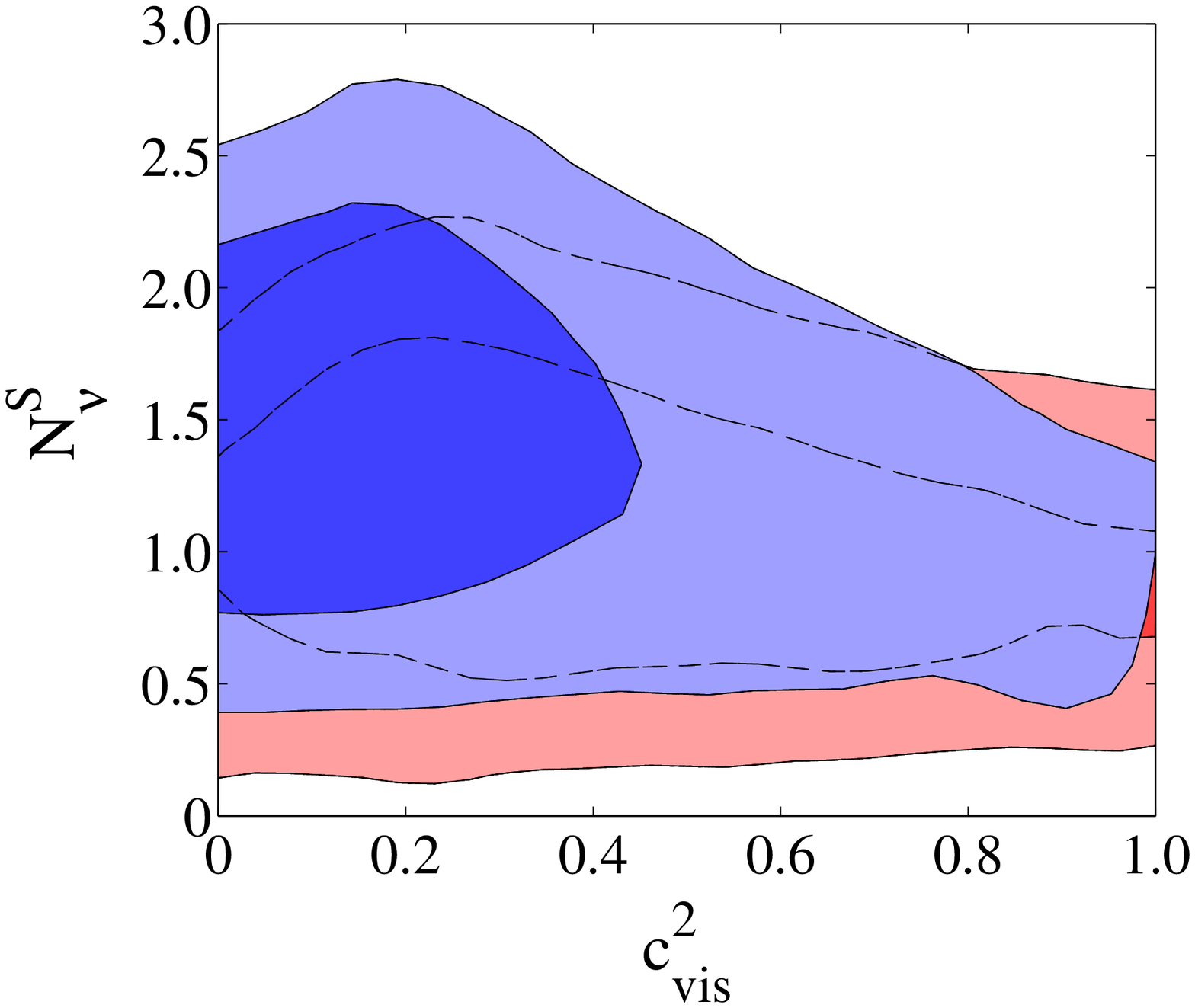}
\caption{$68\%$ and $95\%$ c.l. constraints for the
degeneracy between neutrinos parameters. Red contours refer to model (A) in Table~\ref{delta_n},
while blue contours show model (B).}
\label{ceff-nus}
\end{figure*}

In Table~\ref{delta_n} we report the constraints considering only an excess $\nnus$
in the number of relativistic degrees of freedom over a standard $3$ neutrinos
background.

\begin{table}[h!]
\begin{center}
\begin{tabular}{|l|c|c|}
\hline
\hline
\textbf{Model :} & \textbf{varying $\ceff$, $\cvis$} & \textbf{$\ceff = 1/3$, varying $\cvis$} \\
& \textbf{(A)} & \textbf{(B)} \\
\hline
$\Omega_b h^2$ & $0.02177 \pm 0.00066$ & $0.02262 \pm 0.00049$\\
$\Omega_c h^2$ & $0.135 \pm 0.010$ & $0.143 \pm 0.010$\\
$\tau$ & $0.086 \pm 0.013$ & $0.084 \pm 0.013$\\
$H_0$ & $72.8 \pm 2.1$ & $73.7 \pm 2.2$\\
$n_s$ & $0.989 \pm 0.014$ & $0.978 \pm 0.014$\\
$log(10^{10} A_s)$ & $3.178 \pm 0.035$ & $3.192 \pm 0.035$\\
$A_{SZ}$ & $<1.6$ & $< 1.4$\\
$A_C [{\rm \mu K^2}]$ & $<15.0$ & $< 15.0$\\
$A_P [{\rm \mu K^2}]$ & $<24.8$ & $<24.8$\\
\hline
$\nnus$ & $1.10^{+0.19 +0.79}_{-0.23 -0.72}$ & $1.46^{+0.21 +0.76}_{-0.21 -0.74}$ \\
$\ceff$ & $0.24^{+0.03 +0.08}_{-0.02 -0.13}$ & $1/3$\\
$\cvis$ & $<0.91 $ & $< 0.74$\\
\hline
$\chi^2_{min}$ & $7590.5$ & $7592.0$\\
\hline
\hline
\end{tabular}
\caption{MCMC estimation of the cosmological parameters
considering an extra component $\nnus$ and assuming
a standard background of $3$ relativistic neutrinos.
The perturbation parameters refer to the extra component.
Both $68\%$ and $95\%$ confidence levels for the neutrino parameters are reported.
Upper bounds are at $95 \%$ c.l. .}
\label{delta_n}
\end{center}
\end{table}

As we can see for the results in the table, the evidence for an extra background is solid with
$\nnus=1.46^{+0.76}_{-0.74}$ at $95 \%$ c.l. when only variations in the $\cvis$ component are
considered, while the constraint is $\nnus=1.10^{+0.79}_{-0.72}$ when also variations in
$\ceff$ are considered. Again, the data provide a good determination for $\ceff$ with
$\ceff=0.24^{+0.08}_{-0.13}$ at $95 \%$ c.l., in marginal agreement at about $2 \sigma$ with the
standard $\ceff=1/3$ value. This lower value of $\ceff$, also found in \cite{zahn}, could
hint for a dark radiation component with a varying equation of state, ruling out a
 a massless sterile neutrino. It will be certainly interesting to investigate
if this signal remains in future analyses.  No significant constraint is obtained on $\cvis$.

In Figure~\ref{ceff-nus} we show the degeneracy between the parameters $\nnus$, $\ceff$, and $\cvis$
by plotting the 2D likelihood contours between them. As we can see a degeneracy is present
between $\ceff$ and $\nnus$: models with lower values of $\nnus$ are more compatible with
$\ceff=0$ since the effect of $\ceff$ on the CMB spectrum is smaller.
No apparent degeneracy is present between $\cvis$ and the remaining parameters since
$\cvis$ is weakly constrained by current data.

Since oscillation experiments have clearly established that neutrino are massive, it
is interesting to perform a similar analysis but letting the $3$ neutrino standard
background with $\ceff=\cvis=1/3$ to be massive, and varying the parameter $\Sigma m_{\nu}$ that consider
the sum of masses of the $3$ {\it active} neutrinos. The extra dark radiation
component is assumed massless and we treat the perturbations in it as in the previous sections. 
In Table~\ref{massive} we report the results of this analysis.

\begin{table}[h!]
\begin{center}
\begin{tabular}{|l|c|}
\hline
\hline
$\Omega_b h^2$ & $0.02174 \pm 0.00063$\\
$\Omega_c h^2$ & $0.135 \pm 0.011$\\
$\tau$ & $0.087 \pm 0.014$\\
$H_0$ & $72.7 \pm 2.1$\\
$n_s$ & $0.989 \pm 0.015$\\
$log(10^{10} A_s)$ & $3.179 \pm 0.036$\\
$A_{SZ}$ & $<1.6$\\
$A_C [{\rm \mu K^2}]$ & $<15.9$\\
$A_P [{\rm \mu K^2}]$ & $<26.1$\\
\hline
$\sum m_{\nu}[\rm{eV}]$ & $ < 0.79$ \\
$\nnus$ & $1.12^{+0.21 +0.86}_{-0.26 -0.74}$\\
$\ceff$ & $0.241^{+0.03 +0.09}_{-0.02 -0.12}$\\
$\cvis$ & $<0.92$\\
\hline
$\chi^2_{min}$ & $7590.7$\\
\hline
\hline
\end{tabular}
\caption{MCMC estimation of the cosmological parameters considering $\nnu = 3.04$ massive neutrinos.
Values and 68\% - 95\% errors for the neutrino parameters are reported. Upper bounds are at $95 \%$ c.l. .}
\label{massive}
\end{center}
\end{table}

\begin{figure}[h!]
\includegraphics[scale=0.41]{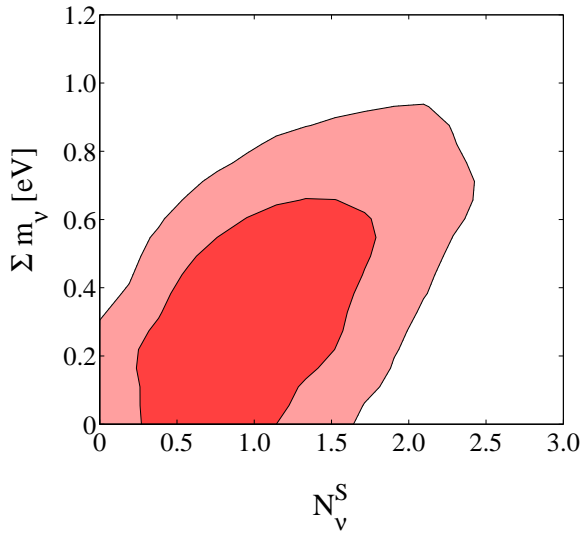}
\caption{Degeneracy in the plane $\sum m_{\nu}-\nnus$ at 68\% and 95\% c.l. .}
\label{sum-nus}
\end{figure}

As we can see, when masses in the active neutrinos are considered, there is
a slightly stronger evidence for the extra background with $\nnus = 1.12^{+0.86}_{-0.74}$.
This is can be explained by the degeneracy present between $\sum m_{\nu}$ and
$\nnus$, well known in the literature (see e.g. \cite{hamann10}) and clearly
shown in Figure~\ref{sum-nus} where we report the 2D marginalized contours in the plane $\sum m_{\nu}- \nnus$.

\subsection{Profile likelihood analysis}

\begin{figure}[h!]
\includegraphics[scale=0.31]{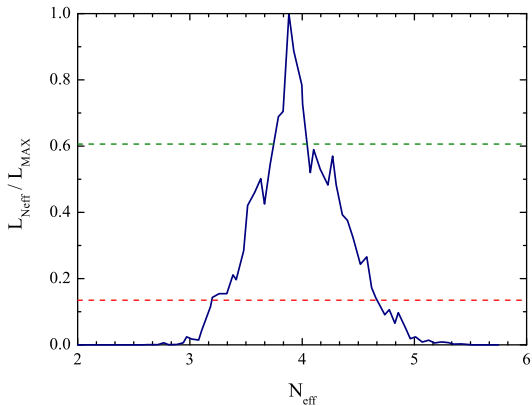}
\caption{Maximum Likelihood ratio $L_{\neff}/ L_{max}$ for $\neff$.
The dashed lines represent the $68 \%$ and $95 \%$ c.l. for a Gaussian
likelihood ($L_{\neff}/ L_{max}=0.6065$ and $L_{\neff}/ L_{max}=0.135$) respectively.}
\label{maxlike}
\end{figure}

Recently, in \cite{verdex}, a model-independent analysis for the extra relativistic
degrees of freedom in cosmological data has been performed claiming no
statistically significant evidence for it.
This simple analysis consists in extracting the maximum likelihood
value $L$ as a function of $\neff$ over the parameter space sampled in the chains,
with a bin width of $0.5$ and constructing a profile likelihood ratio by considering
$ln (L_{\neff}/ L_{max})$ as a function of $\neff$; where $L_{max}$ is the
maximum likelihood in the entire chains.

Here we perform a similar analysis, using however a smaller bin width of
$0.05$ and considering the case where the whole number of relativistic degrees of freedom
$\neff$ is varied while $\cvis=\ceff=1/3$. 
The resulting likelihood ratio $L_{\neff}/ L_{max}$, plotted in Figure \ref{maxlike}, clearly
indicates a preference for a dark radiation component finding that the best fit
model has $\neff=3.88$ with a $\Delta \chi^2=14.56$ respect to the best fit model
with $\neff=3.046$.

We should however point out that the ratio $L_{\neff}/ L_{max}$
presented in Figure \ref{maxlike} is rather noisy. Bayesian methods such as MCMC are 
indeed known to be inaccurate for this purpose (see for example the discussion in \cite{akrami}).
Other methods more appropriate for a frequentist analysis have been 
presented, for example, in \cite{others}.

\section{Conclusions}

In this paper we performed a new search for Dark Radiation, parametrizing it with an effective number of relativistic 
degrees of freedom $\neff$.
We have shown that the cosmological data we considered are clearly suggesting the presence for an extra dark radiation component with $\neff=4.08_{-0.68}^{+0.71}$ at $95 \%$ c.l. . Performing an analysis on its effective sound speed $c_{\rm eff}$ and viscosity $c_{\rm vis}$ parameters, we found  $\ceff=0.312\pm0.026$ and $\cvis=0.29_{-0.16}^{+0.21}$ at $95 \%$ c.l., consistent with the expectations of a relativistic free streaming component ($\ceff$=$\cvis$=$1/3$). Assuming the presence of $3$ standard relativistic neutrinos we constrain the
extra dark radiation component with $\nnus=1.10_{-0.72}^{+0.79}$ and $\ceff=0.24_{-0.13}^{+0.08}$ at $95 \%$ c.l. while $\cvis$ is practically unconstrained. Assuming a mass in the $3$ neutrino component we obtain
further indications for the dark radiation component with $\nnus=1.12_{-0.74}^{+0.86}$ at $95 \%$ c.l. .
From these results we conclude that Dark Radiation currently represents
one of the most relevant anomaly for the $\Lambda$-CDM scenario. 

When comparison is possible, our results are in good agreement with the most recent analysis
presented in \cite{zahn} that uses a different choice of datasets (for example, we don't consider 
matter fluctuations data from Lyman-$\alpha$ as in \cite{zahn}) and an independent 
analysis method.

Dark Radiation will be severely constrained in the very near future by the Planck satellite data, where a precision on $\neff$ of about $\Delta \neff \sim 0.2$ is expected (see e.g. \cite{galli} and \cite{keating}) only from CMB data.
\\

\section{Acknowledgments}
We thank Ryan Keisler for providing us with the likelihood code for the SPT data.
We thank Luca Pagano for help.  This work is supported by PRIN-INAF, "Astronomy probes fundamental physics".
Support was given by the Italian Space Agency through the ASI contracts Euclid- IC (I/031/10/0).

\end{document}